\documentclass[12pt,a4paper]{article}
    \usepackage[format=hang]{caption}
    \usepackage{epsfig,amssymb,amsmath,graphicx,subcaption,verbatim,hyperref,ulem,bm,float,tensor}
    \usepackage{xcolor}
    \usepackage{color}
    \usepackage{tikz}
    \usetikzlibrary{decorations.markings}
    \usepackage{amsthm} 
    \usepackage{physics}
    \usepackage{booktabs} % for professional-looking tables

    \definecolor{BrickRed}{RGB}{203, 65, 84}
    \definecolor{Violet}{HTML}{EE82EE}
    \definecolor{OliveGreen}{HTML}{556B2F}
    \definecolor{RoyalBlue}{RGB}{25,41,88}
    
    \hypersetup{colorlinks=true, linkcolor=BrickRed, citecolor=Violet, filecolor=OliveGreen, urlcolor=RoyalBlue, filebordercolor={.8 .8 1}, urlbordercolor={.8 .8 0}}
    \newcommand{\be}{\begin{equation}}
    \newcommand{\ee}{\end{equation}}

\begin{document}

\title{
        {\bf {\large {Exploring the phase diagram of $SU(2)_4$ strange correlator}}}\\[20pt]}
\author{ {Ce Shen}\\[20pt]
    Beijing Institute of Mathematical Sciences and Applications\\
}

\date{\today}
\maketitle
\vskip 20pt

\begin{abstract}
    We investigate the phase diagram of a quantum many-body system constructed via the strange correlator approach, based on the non-Abelian $SU(2)_4$ fusion category, to probe topological phase transitions. Using tensor network methods, we numerically compute the half-infinite chain entanglement entropy derived from the dominant eigenvector of the transfer matrix and map the entropy across a spherical two-dimensional parameter space. Our results reveal a phase diagram significantly more complex than previously reported\cite{cft_from_tqft}, including a gapless phase consistent with a conformal field theory (CFT) of central charge $c=1$. Critical lines separating distinct phases are identified, with one such line—bounding the CFT phase—exhibiting a higher central charge $c=2$, indicative of an unconventional critical regime. 
\end{abstract}

\vfill

%PACS:
\newpage

\section{Introduction}
% briefly introduce strange correlator, its connection with anyon chain, and the motivation of this work
Understanding how a topologically ordered system can yield a conformal field theory(CFT) is one of the most exciting challenges in modern theoretical physics. In topologically ordered systems, e.g. certain quantum Hall states and spin liquids, the ground state exhibits topological long-range entanglement that is not characterized by any local order parameter.\cite{Chen_Gu_Wen_2010_LU,Wen_bosonic_TO,Wen_fermionic_TO,ZHWang_2012_TO_by_EE} These phases are robust against local perturbations and can host exotic excitations called anyons, which obey non-standard (often non-Abelian) exchange statistics.\cite{wilczek1982anyon,Kitaev2003Fault,Kitaev2005Anyons,Nayak2007Non-Abelian} Conformal field theories (CFTs), on the other hand, describe critical systems at a continuous phase transition where scale invariance and conformal symmetry dictate the behavior of correlations at all length scales.\cite{BPZ1984CFT,MooreSeiberg1989CFT,francesco2012CFT}

Our work builds on a rich history of research that bridges topological phases with critical behavior. Early studies on interacting anyon chains, such as the seminal ``golden chain'' model introduced by \cite{KitaevWang2007GoldenChain}, revealed that chains of Fibonacci anyons can exhibit critical behavior governed by a CFT. Subsequent investigations, including those on parafermionic models \cite{Fendley2004TO_critical,fendley_2020}, have further underscored the connection between anyonic systems and conformal criticality.

A particularly fruitful avenue for linking these two domains is the concept of the “strange correlator.” Although the foundational ideas connecting topological order and criticality appeared in earlier works\cite{KitaevWang2007GoldenChain,Fendley2004TO_critical}, the term “strange correlator” was introduced by You et al.\cite{you2014SC} in the context of symmetry-protected topological (SPT) phases. In its later reinterpretation, the strange correlator is defined as the inner product between a topologically ordered state and a simple product state. This reinterpretation has been pivotal; subsequent studies have extended the framework to establish a systematic correspondence between fully topological models and CFTs \cite{Frank2018mapping_topo_to_CFT,cft_from_tqft,Hung2023exact_fixed_point,zeng2023virasoro_generator,wang2022virasoro}.

% A remarkable and fruitful approach to bridge these two seemingly different worlds is the concept of the ``strange correlator''. 
% It should be noted that while the foundational ideas connecting topological phases with critical theories were established in early studies\cite{KitaevWang2007GoldenChain,Fendley2004TO_critical}, the specific terminology ``strange correlator'' originated with \cite{you2014SC} and was later reinterpreted.
% You et al. \cite{you2014SC} introduced the term ``strange correlator'' in the context of SPT phases that differed from its later re-interpretation as the inner product between a Levin-Wen ground state and a simple product state. Subsequent work \cite{Frank2018mapping_topo_to_CFT,cft_from_tqft,Hung2023exact_fixed_point,zeng2023virasoro_generator,wang2022virasoro} extended this framework to establish a connection between fully topological order models and CFTs.

In the strange correlator construction one considers the inner product between a topologically ordered state $\ket{\psi}$ and a product state $\ket{\Omega}$. The topologically ordered state $\ket{\psi}$ is taken to be the ground state of a 2+1-dimensional Levin-Wen string-net model\cite{Vidal2008TN_rep_LW,Gu2009TN_rep_LW}, which is defined by local tensors whose indices are contracted according to the fusion rules of an input fusion category.
The inner product $\bra{\psi}\ket{\Omega}$ represents a two-dimensional partition function that can be interpreted as describing some classical statistical mechanical system, where correlations between degrees of freedom are governed by statistical weights of the system. Tuning parameters in the chosen product state then drives the system across different phases, effectively interpolating between gapped phases and cross a critical phase described by a CFT\cite{cft_from_tqft}.

In our work we study this strange correlator construction with an input fusion category $SU(2)_4$. This theory is particularly appealing because it hosts a richer spectrum of anyon types than, for example, the Ising or Fibonacci theories\cite{Frank2018mapping_topo_to_CFT}, yet it remains tractable. 
% also cite Zhenghan Wang's su(2)_4 encoding quantum info
By choosing the Levin-Wen model with $SU(2)_4$ as input and tuning the parameters of the product state, our numerical investigations reveal a two-dimensional phase diagram. In certain regions the effective classical partition function is gapped and retains topological features, while in other regions it exhibits scale invariance and universal behavior described by a CFT. 
% Notably, our analysis uncovers several phase transition points between different topological sectors, thus highlighting the subtle interplay between topological protection and critical fluctuations.

We use entanglement entropy as a probe to study these phases. The partition function obtained from the strange correlator can be represented as a tensor network\cite{Frank2018mapping_topo_to_CFT,Frank2022haagerup} whose transfer matrix encodes the physical properties of the system. We then find the dominant eigenstate of this transfer matrix using the VUMPS algorithm\cite{zauner2018VUMPS,TensorKit,MPSKit} based on infinite matrix product states (iMPS). The entanglement entropy of the left and right half-infinite chain of this dominant eigenstate is computed numerically. At a critical point, this entropy scales logarithmically with effective correlation length, with the scaling coefficient directly related to the central charge of the underlying CFT\cite{Cardy2004EE_and_QFT,Cardy2009EE_and_CFT,Tagliacozzo2008Entanglement_scaling,Pollmann2009Entanglement_scaling,Frank2015Entanglement_scaling}. Conversely, in a gapped phase the entanglement saturates to a constant value (the ``area law''\cite{Eisert2010RMP_area_law,Pollmann2017SPT}), thereby providing a clear numerical signature of the phase.

% By applying the strange correlator framework to a string-net with $SU(2)_4$ input category, we extend these ideas and offer a complementary perspective on how topological order and criticality can coexist and be tuned continuously within a single model.

The remainder of the paper is organized as follows. In Section 2 we detail the construction of the tensor network representation for the Levin-Wen ground state with an input fusion category $SU(2)_4$, and describe how the strange correlator is obtained from this representation. We also discuss the formulation of the transfer matrix and the entanglement analysis that underpins our phase diagram. Section 3 presents our numerical results, including the phase diagram on the sphere, a detailed discussion of several critical points (where isolated CFT behaviors emerge), entire phase transition lines along which every point exhibits conformal invariance, and an extended CFT phase. Finally, in Section 4 we summarize our conclusions, highlight the interplay between topological order and criticality revealed by our study, and discuss potential avenues for future research.

\section{Strange correlator and tensor network representation}
% In this section, we would like to review the strange correlator construction and its tensor network representation.
\subsection{Tensor network representation of Levin-Wen ground state}
The Levin-Wen model is a microscopic Hamiltonian framework formulated to capture a broad class of topological phases in two-dimensional quantum systems. The model is constructed on a trivalent lattice, and takes as input a fusion category, which encodes the algebraic data of anyonic excitations through fusion rules, quantum dimensions, and the F-symbols. The local Hilbert space on each edge of the lattice is spanned by the simple objects of the fusion category, and the Hamiltonian is defined by a set of local projectors that enforce the fusion constraints of the category.

The tensor network representation of the ground state in the Levin-Wen model\cite{Vidal2008TN_rep_LW,Gu2009TN_rep_LW} provides a powerful framework for capturing and analyzing the rich topological phases present in these systems. By expressing the ground state as a network of local tensors that are subject to specific fusion constraints, one can map the nonlocal properties of the model onto a structure amenable to both analytical and numerical investigations. Importantly, this formulation allows us to harness an extensive suite of tensor network algorithms\cite{Orus2018TN_review,Vidal2015TNR,Yang2017loopTNR,zauner2018VUMPS} to study various properties of the system.

\begin{figure}[h]
    \centering
    \begin{subfigure}{0.48\textwidth}
        \centering
        \includegraphics[width=\textwidth]{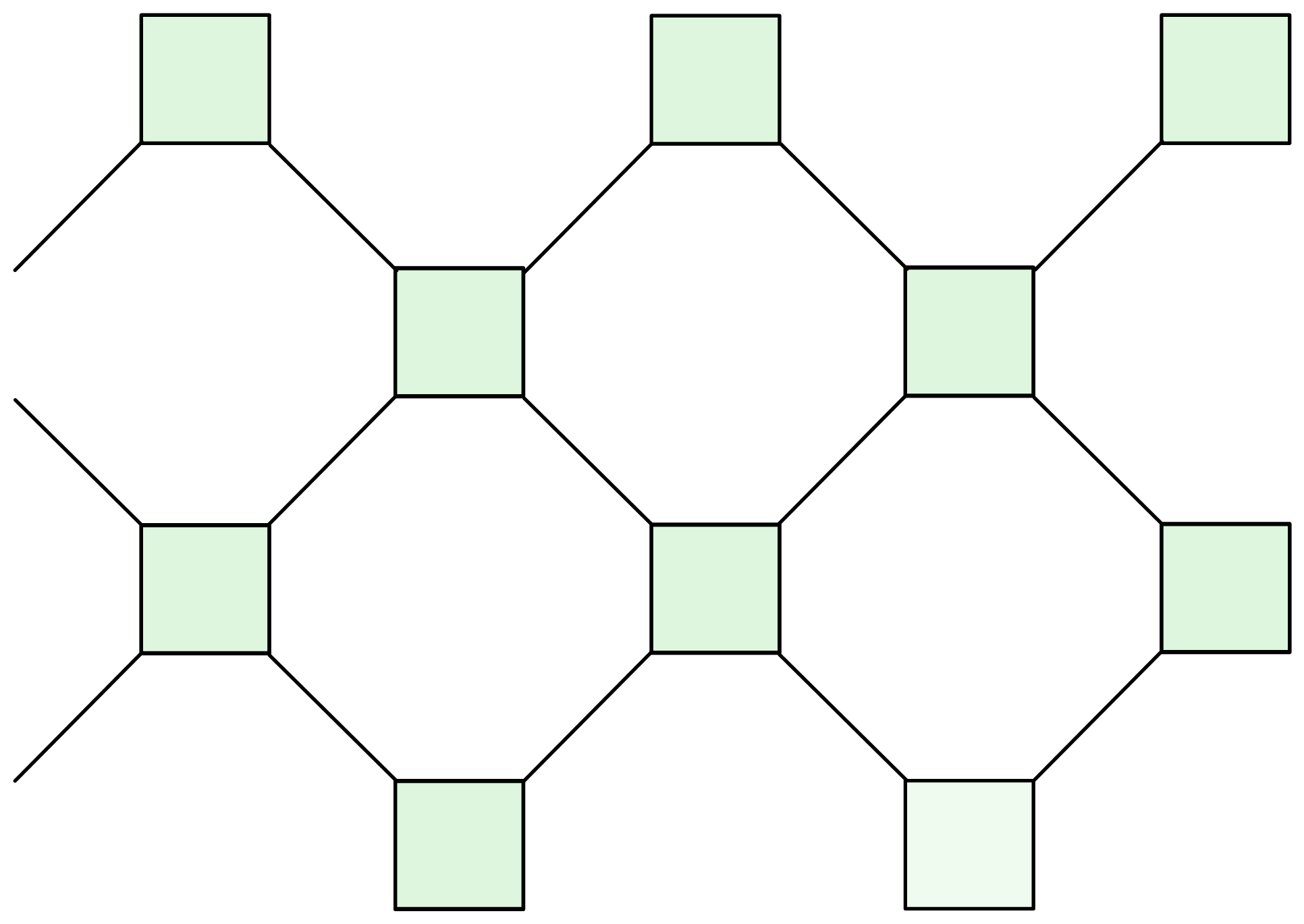}
        \caption{}
        \label{fig:lw_gs_1}
    \end{subfigure}
    \hfill
    \begin{subfigure}{0.4\textwidth}
        \centering
        \includegraphics[width=\textwidth]{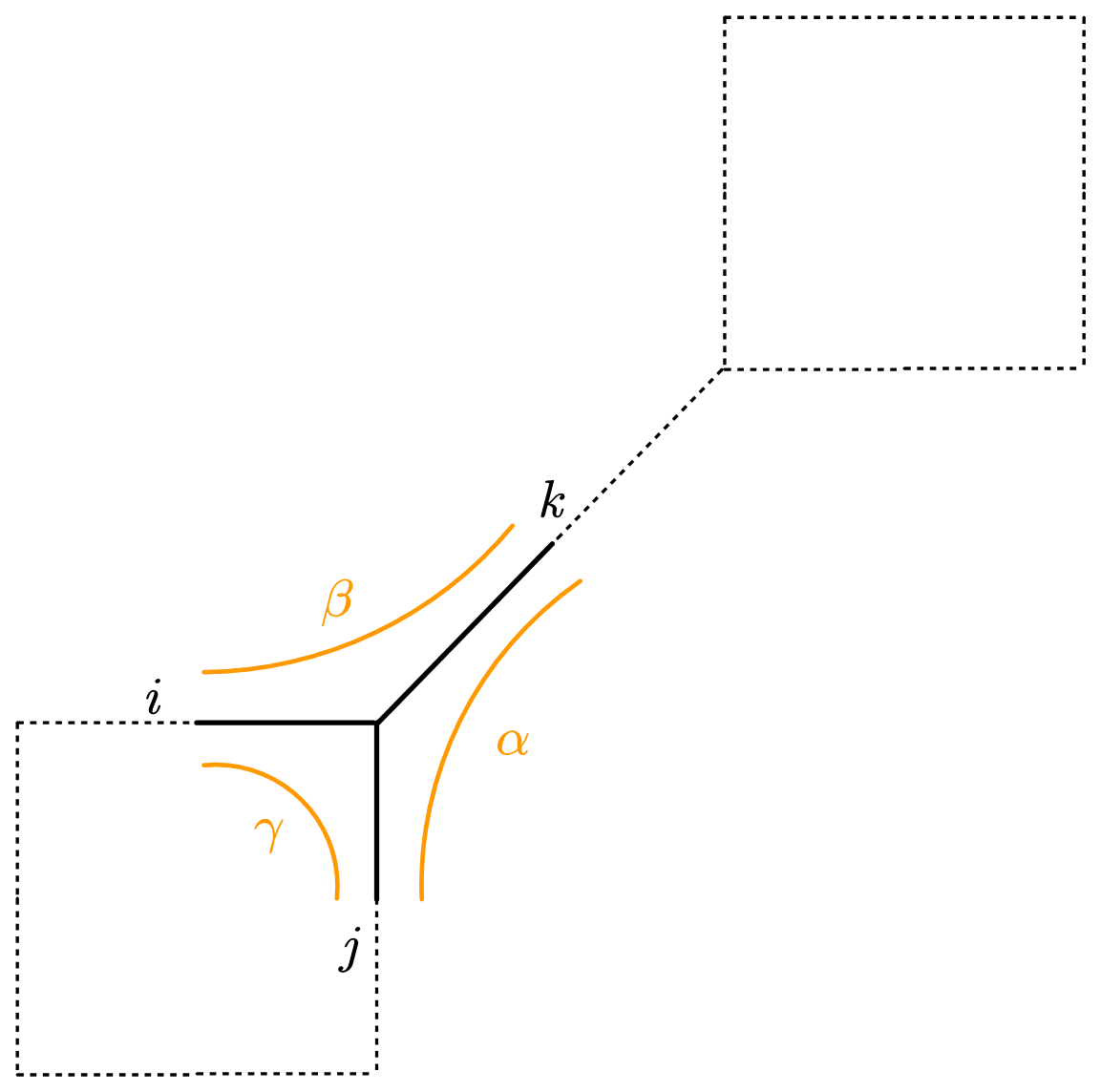}
        \caption{}
        \label{fig:lw_gs_2}
    \end{subfigure}
    \caption{(a) The square-octagon lattice. (b) The tensor network representation of the ground state of a Levin-Wen model on square-octagon lattice. The whole tensor network is contracted from the 6-legged tensor $T^{ijk}_{\alpha\beta\gamma}$ shown here.}
    \label{fig:lw_gs}
\end{figure}

Through out this paper, we will consider the Levin-Wen model on the square-octagon lattice, as shown in Fig.\ref{fig:lw_gs_1}. For convenience we will refer to the edge of the squares as ``square edge'' and the edge shared by neighboring octagons as ``octagon edge''.

The ground state of the Levin-Wen can be represented by a tensor network:
\begin{equation}
    \ket{\psi} = \sum_{ijk} \prod T^{ijk}_{\alpha\beta\gamma} \ket{ijk}
\end{equation}
where we use $\prod T^{ijk}_{\alpha\beta\gamma}$ to represent the contraction of a network of 6-legged $T$-tensors, and $\ket{ijk}$ to represent a basis state in the Levin-Wen Hilbert space with square edges and octagon edges fixed to the labels $i$, $j$ and $k$ etc.
For the square-octagon lattice, the $T$-tensor is shown in Fig.\ref{fig:lw_gs_2} where $i$ and $j$ are the indices of the square edge, and $k$ is the index of the octagon edge. $\alpha$, $\beta$ and $\gamma$ are the ancillary indices necessary to decompose the possibly long-range entangled state $\ket{\psi}$ into contraction of local tensors. In our setup of square-octagon lattice, the values of the $T$-tensors are, up to a global constant, given by\cite{Frank2018mapping_topo_to_CFT,zeng2023virasoro_generator}:
\begin{equation}
    T^{ijk}_{\alpha\beta\gamma} = \left(d_\alpha^{1 / 8} d_\beta^{1 / 8} d_\gamma^{1 / 4}\right) \left(d_i d_j d_k\right)^{-\frac{1}{4}}\left(d_\alpha d_\beta d_\gamma\right)^{-\frac{1}{2}} \sqrt{d_k d_i d_{\gamma} d_{\alpha}} [F^{ik\alpha}_{\gamma}]_{j\beta}
\end{equation}
where $d_x$ is the quantum dimension of the simple object $x$, and $F$ is the F-symbol of the input fusion category.

The $SU(2)_4$ fusion category has 5 simple objects, labeled by $0$, $1$, $2$, $3$ and $4$. The quantum dimensions are $d_0=d_4=1$, $d_1=d_3=\sqrt{3}$, $d_2=2$. For the fusion rules and F-symbols, we refer the readers to the appendix \ref{app:su2_4}.

\subsection{Tensor network representation of strange correlator}
\begin{figure}[h]
    \centering
    \begin{subfigure}{0.4\textwidth}
        \centering
        \includegraphics[width=\textwidth]{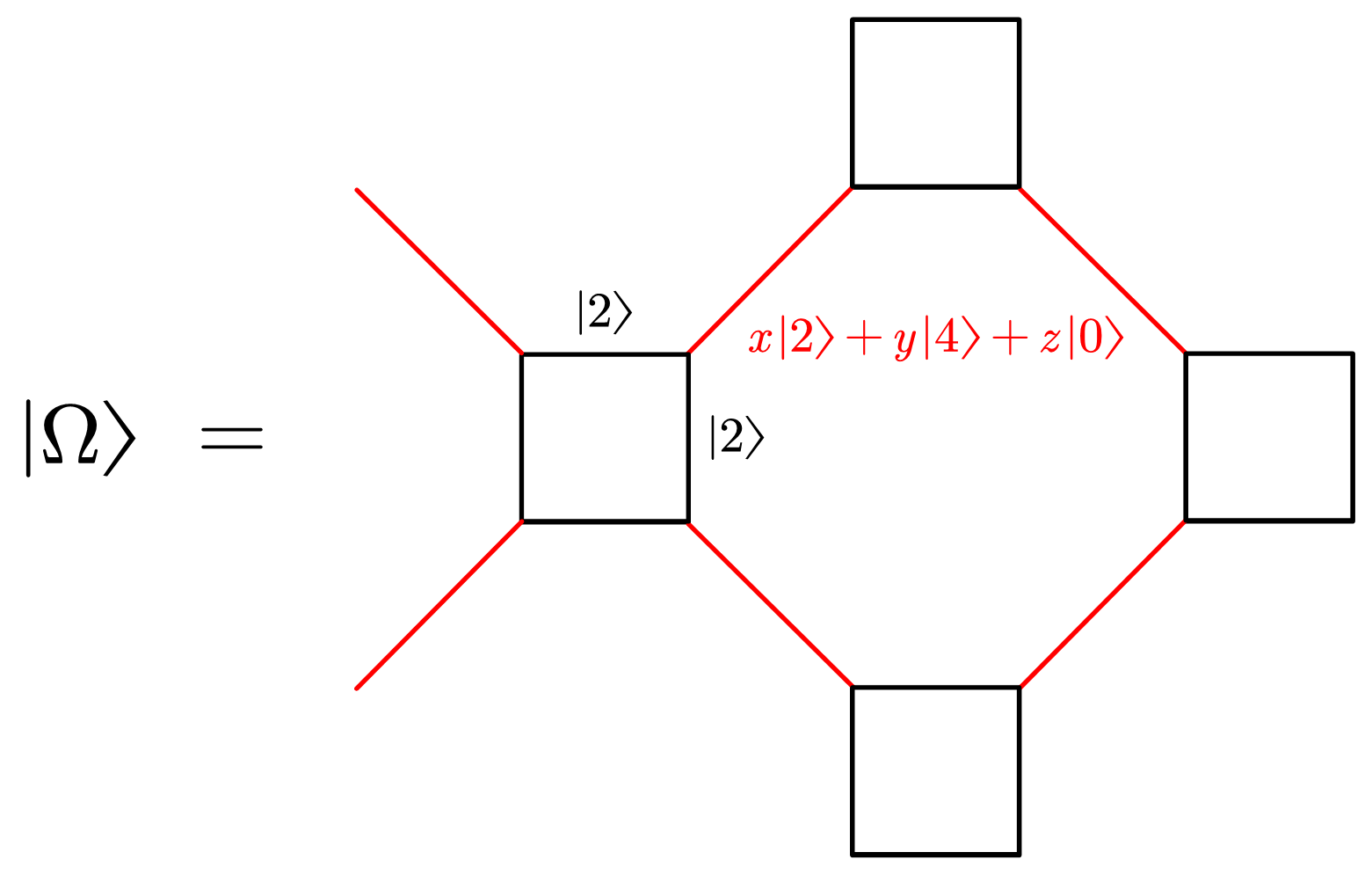}
        \caption{}
        \label{fig:sc_1}
    \end{subfigure}
    \hfill
    \begin{subfigure}{0.5\textwidth}
        \centering
        \includegraphics[width=\textwidth]{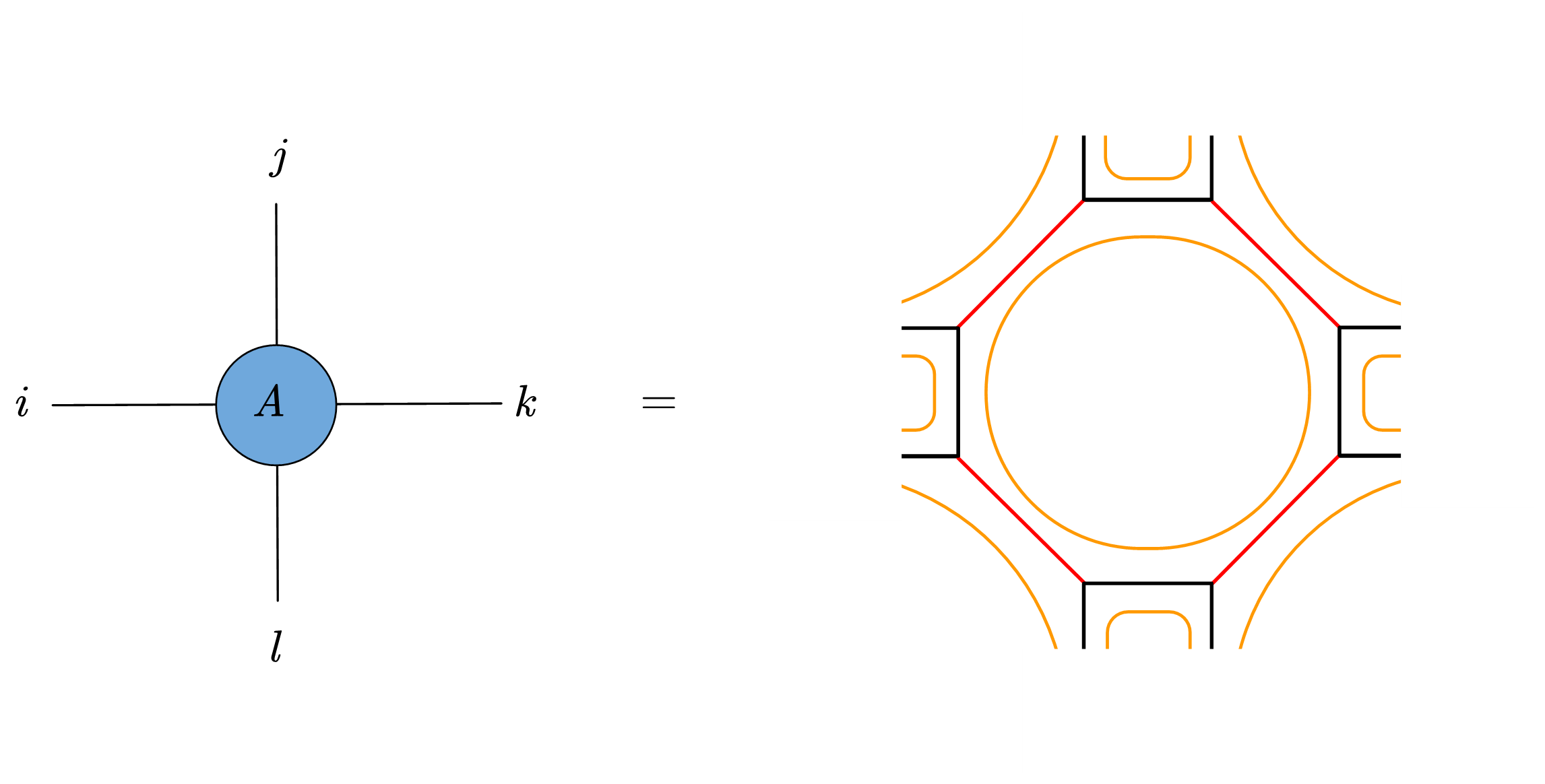}
        \caption{}
        \label{fig:sc_2}
    \end{subfigure}
    \caption{(a) The product state $\ket{\Omega(x,y,z)}$. (b) The tensor network representation of the inner product $\bra{\psi}\ket{\Omega(x,y,z)}$. The local tensors are contracted in this specific manner to make a network of $A_{ijkl}$ tensors in square lattice.}
    \label{fig:sc}
\end{figure}

For the product state $\ket{\Omega}$, we consider the configuration as shown in Fig.\ref{fig:sc_1}. 
The state on each square edge is fixed to $\ket{2}$ in the local Hilbert space, while the state on each octagon edge is taken to be $x\ket{2} + y\ket{4} + z\ket{0}$. We impose the normalization condition $x^2+y^2+z^2=1$,as an overall scaling factor does not affect the system. This combination is taken to respect the fusion rule $2\otimes 2 = 0+2+4$ in the $SU(2)_4$ fusion category.

The strange correlator is obtained by taking inner product $\bra{\psi}\ket{\Omega(x,y,z)}$. Its tensor network representation is shown in Fig.\ref{fig:sc_2}. The local tensors are contracted in this specific manner to re-express the strange correlator as a square network of $A_{ijkl}$ tensors. In the following we may use $x\ket{2} + y\ket{4} + z\ket{0}$ to denote either the product state $\ket{\Omega(x,y,z)}$ or the corresponding strange correlator $\bra{\psi}\ket{\Omega(x,y,z)}$ when the meaning is clear from context.

% \subsection{Notes on entropy calculation}

\section{The phase diagram}
In this section we give a detailed description of the phase diagram of the $SU(2)_4$ strange correlator and provide numerical evidence for the identification of several critical points/lines/regions. 

We continuously change the strange correlator by tuning the three parameters $x$, $y$ and $z$ in the product state $\ket{\Omega(x,y,z)}$.
The transfer matrix of this model is obtained by contracting a row of the tensor $A_{ijkl}(x,y,z)$ shown in Fig.\ref{fig:sc_2}.
In thermodynamic limit, the transfer matrix is represented as an infinite matrix product operator (iMPO) and its dominant eigenvector can be obtained using iMPS methods. Numerically, we use VUMPS algorithm\cite{zauner2018VUMPS} and Julia's ``MPSKit'' package\cite{MPSKit,TensorKit} to obtain the dominant eigenvector of the transfer matrix. The entanglement entropy of the dominant eigenvector of the transfer matrix is then calculated for the half-infinite chain.

% spherical plot
\subsection{Spherical plot of entanglement entropy}
This model has a two dimensional spherical parameter space under the normalization $x^2+y^2+z^2=1$. For convenience, we also use the spherical coordinates $\theta$ and $\phi$ to describe the parameter space. The two parametrizations are related by $x=\sin\theta\cos\phi$, $y=\sin\theta\sin\phi$, $z=\cos\theta$. In certain cases we also use the normalization condition $z=1$ and projectthe phase diagram onto the $xOy$ plane.

\begin{figure}[h]
    \centering
    \includegraphics[width=\textwidth]{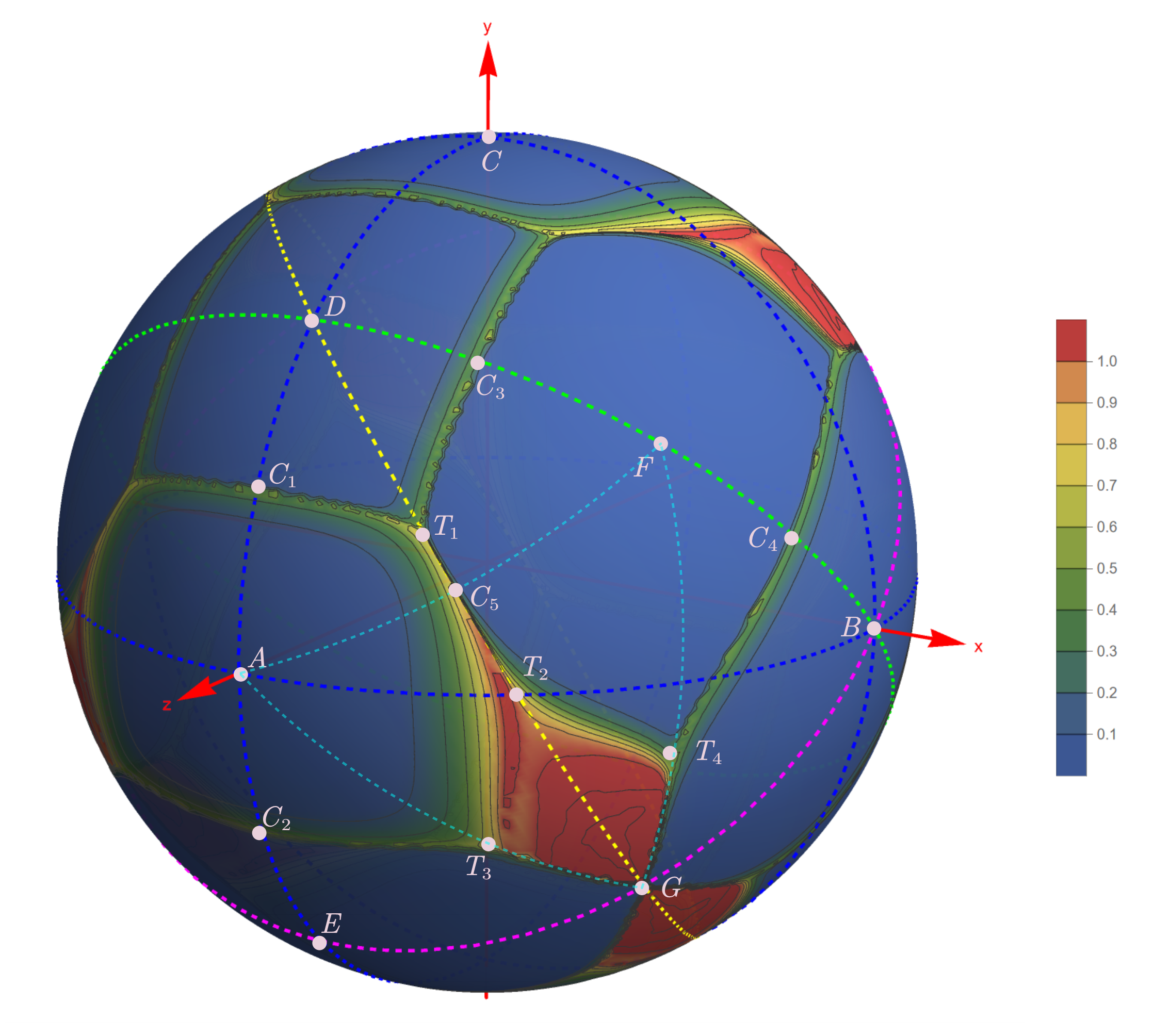}
    \caption{The spherical contour plot of half-infinite chain entanglement entropy of $SU(2)_4$ strange correlator. This plot is obtained by setting the iMPS bond dimension $D=20$.}
    \label{fig:spherical_plot}
\end{figure}

Fig.\ref{fig:spherical_plot} illustrates the phase diagram of the $SU(2)_4$ strange correlator as determined by the spherical contour plot of entanglement entropy. The color of the point represents the value of the entanglement entropy. 
The phase diagram is divided into several phases. Certain phase transition lines between gapped phases are clearly visible, indicating sharp changes in the entanglement entropy and hence in the underlying strange correlator.
\begin{itemize}
    \item Gapped phases: The gapped phases are characterized by near-zero entropy, which appear as blue regions in the plot. In Fig.\ref{fig:spherical_plot}, the blue regions centered at $A,B,C,D,E,F$ are all gapped phases. With a slight abuse of notation, we use $X$ to denote the gapped phase centered at $X$, where the meaning should be clear from context. The gapped phases $A,D,F$ are exactly the type I, II and III gapped phases identified in \cite{cft_from_tqft}, which correspond to the Frobenius algebras $\mathcal{A}_{I}=0$, $\mathcal{A}_{II}=0+4$ and $\mathcal{A}_{III}=0+2+4$ respectively. The gapped phases $B,C,E$ are new gapped phases that are not identified in \cite{cft_from_tqft}.
    \item Gapless phases: The gapless phases are characterized by nearly maximal entropy due to long-range entanglement, and appear as red regions in the plot. The red region enclosed roughly by $T_2, T_3, T_4, G$ is a gapless phase that is not identified in \cite{cft_from_tqft}. We later give numerical evidence that this red region is a CFT phase with central charge $c=1$.
\end{itemize}

% description of the points
% Reference \cite{cft_from_tqft} supposed that the line $T_1 T_2$ represents a phase transition line between the gapped phases $A$ and $F$, with the point $T_1$ identified as a tricritical point separating the gapped phases $A,D$ and $F$. However, the entanglement entropy analysis in the current paper indicates that the line $T_1T_2$ possesses a small but nonzero width and lies within the CFT phase (depicted as the red region in the plot). The current numerical results suggest that the CFT phase extends as a tapering tail converging towards $T_1$, while predominantly occupying the region enclosed by $T_2, T_3, T_4$ and $G$. We suppose that $T_1$ is not a tricritical point but rather a tetracritical point between the three gapped phases $A,D,F$ and the $c=1$ CFT phase.

We now give a comprehensive description of the position of various points and lines in the phase diagram.\footnote{The position of the points $C_1,C_3$ and $C_5$ can be calculated from the perspective of anyon condensation or Kramers-Wannier duality\cite{our_new_paper}.}
\begin{itemize}
    \item All dashed lines in  Fig.\ref{fig:spherical_plot} are (part of) big circles on the unit sphere. 
    \item $A,B,C,D,E,F$: These points are gapped and completely disentangled, the entanglement entropy at these points are exactly zero. $A,B,C$ are the intersection points of the $z,x,y$ axes and the unit sphere. $D$ lies at the midpoint of the arc $AC$, and $F$ lies at the midpoint of the arc $DB$. The points $D,E,F$ correspond to states $\ket{0}+\ket{4}$, $\ket{0}-\ket{4}$ and $\ket{0}+\sqrt{2}\ket{2}+\ket{4}$ respectively. The blue, green and magenta dashed lines are the big circles on the unit sphere that pass through these points.
    \item $C_1,C_2,C_3,C_4$: These points are critical points on the phase transition lines between gapped phases. The entanglement entropy at these points are non-zero and scale logarithmically with the effective correlation length of iMPS. The four phase transition lines on which these points lie are all of Ising type and have central charge $c=1/2$. The critical points $C_1,C_2,C_3,C_4$ are located at the quadrisection points of the arcs $DE$ and $DB$ respectively. At these points, the phase transition lines are orthogonal to the dashed lines that interpolate neighboring gapped phases. The coordinates of these points are easily calculated from this fact. For example, under the normalization condition $z=1$ the points $C_1$ and $C_3$ have coordinates $(x,y) = (0,\sqrt{2}-1)$ and $(x,y) = (2-\sqrt{2},1)$ respectively. The spherical coordinates of $C_1$ and $C_2$ are $(\theta,\phi)=(\pi/8,\pm\pi/2)$.
    \item $C_5$: This point is the intersection of $AF$ and $DG$, and it lies at the midpoint of the arc $AF$. Its spherical coordinates are $(\theta,\phi)=(\pi/6,\arctan(1/\sqrt{2}))$. Current numerical analysis suggests that at this point the strange correlator represents a CFT point with central charge $c=1$.
    % \item $T_1$: This point is supposed to be the tetracritical point between the three gapped phases $A,D,F$ and the $c=1$ CFT phase. It is on the KW line (the yellow dashed line) but its precise position is currently unknown. Our numerical analysis suggests the length of the arc $DT_1$ is between $0.58$ and $0.6$. Alternatively, under the $z=1$ normalization this point's $x$-coordinate is in the range $(0.45,0.46)$ and its $y$-coordinate is determined by the equation of the KW line $y=-\sqrt{2}x+1$.
    \item $T_1$: This point is the tricritical point between the three gapped phases $A,D,F$. It is on the KW line (the yellow dashed line $DG$) but its precise position is currently unclear. Our numerical analysis suggests the length of the arc $DT_1$ is between $0.58$ and $0.6$. Alternatively, under the $z=1$ normalization this point's $x$-coordinate is in the range $0.45< x(T_1) <0.46$ and its $y$-coordinate is determined by the equation of the KW line $y=-\sqrt{2}x+1$.
    \item $T_2$: This point is the intersection of the KW line and the line $y=0$. The length of the arc $AT_2$ is $\arctan(1/\sqrt{2})$, and the length of arcs $DT_2$ and $T_2B$ are both $\arctan(\sqrt{2})$. This point also represents a $c=1$ CFT.
    \item $T_1T_2$: This line is the phase transition line between the gapped phases $A$ and $F$. The length of the arc $T_1T_2$ is $\arccot(2\sqrt{2})$. We later give numerical evidence that the central charge of this phase transition line is $c=1$.
    \item $G$: This point is the intersection point of four phases - two gapped phases $E,B$ and two CFT phases. It lies at the intersection of four dashed lines: the line $DG$, the line $EB$, the line $AG$ and the line $GF$. The point $G$ is the midpoint of the arc $EB$. Under the $z=1$ normalization this point has coordinate $(x,y) = (\sqrt{2},-1)$. Entanglement scaling at this point shows that it is not a critical point.
    \item $T_3, T_4$: These points are tricritical points between the CFT phase and two gapped phases. $T_3$ and $T_4$ are on the lines $AG$ and $AF$ respectively, but their precise positions are currently unknown. Later we give estimation of the $x$-coordinate of $T_3$ and the $y$-coordinate of $T_4$ under the $z=1$ normalization.
    \item $GT_3$ and $GT_4$: They are the phase transition lines between the CFT phase and the gapped phases $E$ and $B$ respectively. The central charge of these phase transition lines are $c=2$. Under the $z=1$ normalization, the line $GA$ has equation $y=-1/\sqrt{2}x$ and the line $GF$ has equation $x=\sqrt{2}$.
\end{itemize}

\subsection{Symmetry}
One of the benefits of plotting the phase diagram on the sphere is that it makes the symmetry of the phase diagram more apparent. Let $O$ denote the origin point of $xyz$ coordinate system. The phase diagram is symmetric under each of the following three reflections:
\begin{itemize}
    \item $x\to -x$: This is a reflection about the plane $AOC$.
    \item $z\leftrightarrow y$: This is a reflection about the plane $BOD$.
    \item $z\leftrightarrow -y$: This is a reflection about the plane $BOE$.
\end{itemize}

\begin{figure}[H]
    \centering
    \begin{subfigure}{0.45\textwidth}
        \centering
        \includegraphics[width=\textwidth]{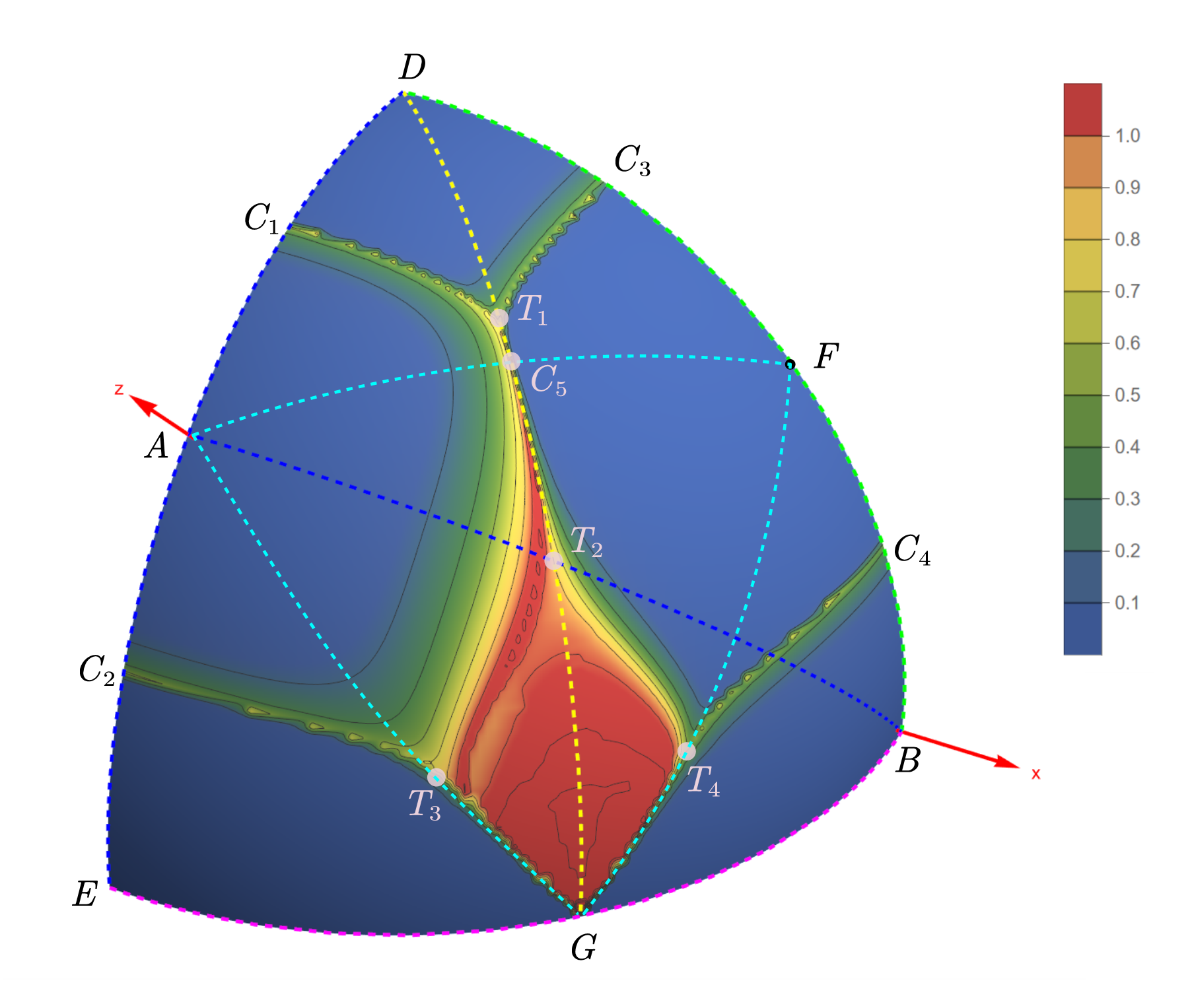}
        \caption{}
        \label{fig:triangular_sphere}
    \end{subfigure}
    \hfill
    \begin{subfigure}{0.45\textwidth}
        \centering
        \includegraphics[width=\textwidth]{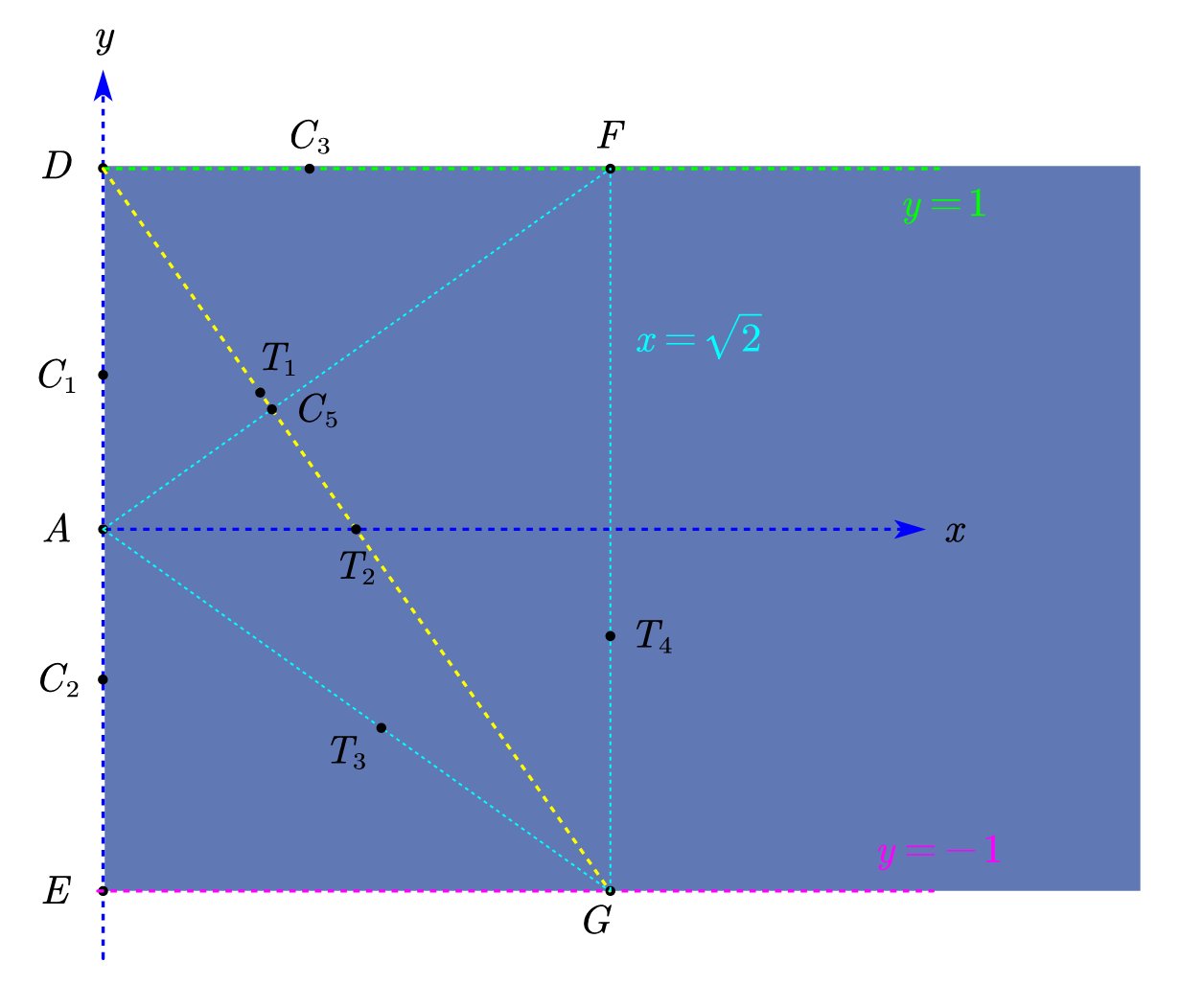}
        \caption{}
        \label{fig:rectangular_sphere}
    \end{subfigure}
    \caption{(a) The equilateral right triangle occupies one eighth of the unit sphere. The whole phase diagram can be obtained by applying reflections to this region. (b) The same region as in (a), plotted on the $xOy$ plane.}
    % \label{fig:DE}
\end{figure}

The phase diagram is symmetric under the three reflections, so the phase diagram is effectively divided into 8 equivalent regions. We take a spherical equilateral right triangle $\triangle DEB$, as depicted in Fig.\ref{fig:triangular_sphere}. All other regions of the phase diagram can be obtained by applying the above reflections to this region. The spherical angles $\angle DEB = \angle EBD = \angle BDE = \pi/2$. The points $A,G,F$ are the midpoints of the arcs $DE, EB$ and $BD$ respectively. The spherical triangle $\triangle AGF$ is also equilateral and has angles $\angle AGF = \angle GFA = \angle FAG = 2\arctan(1/\sqrt{2})$.

The symmetries are not independent. The combined application of all the three reflections brings the point $(x,y,z)$ to its antipodal point $(-x,-y,-z)$, which is automatically a symmetry of the system. From our construction of the $A_{ijkl}$ tensor, the antipodal transformation leaves the tensor $A_{ijkl}$ invariant because it contains an even number of octagon edges.

% #TODO: talk about the symmetry by KW plane???

\subsection{The critical points and lines}
We now plot the entanglement entropy along several interesting lines on the diagram and show the entanglement scaling results of several critical points/lines.
For convenience, we re-plot the region of phase diagram represented by Fig.\ref{fig:triangular_sphere} on the $xOy$ plane under the normalization condition $z=1$, as shown in Fig.\ref{fig:rectangular_sphere}.

\subsubsection{Along the line $DE$ and $DB$}
\begin{figure}[h]
    \centering
    \begin{subfigure}{0.45\textwidth}
        \centering
        \includegraphics[width=\textwidth]{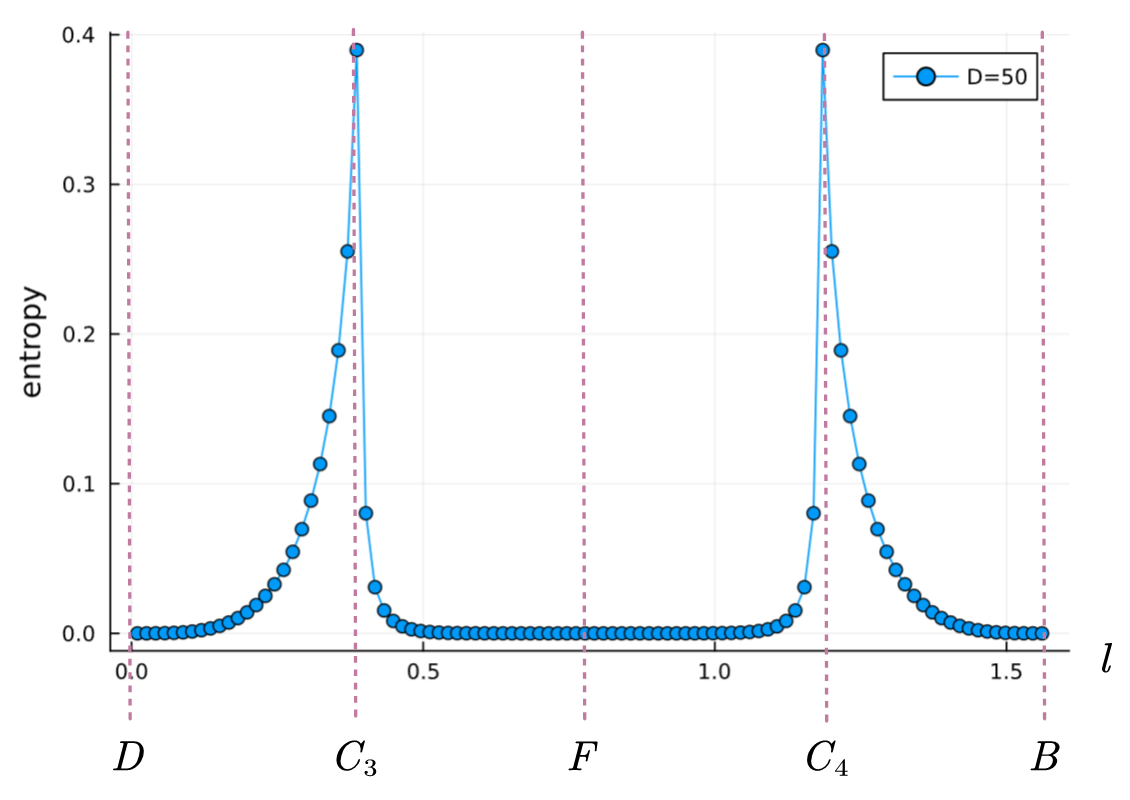}
        \caption{}
        \label{fig:DE_line}
    \end{subfigure}
    \hfill
    \begin{subfigure}{0.45\textwidth}
        \centering
        \includegraphics[width=\textwidth]{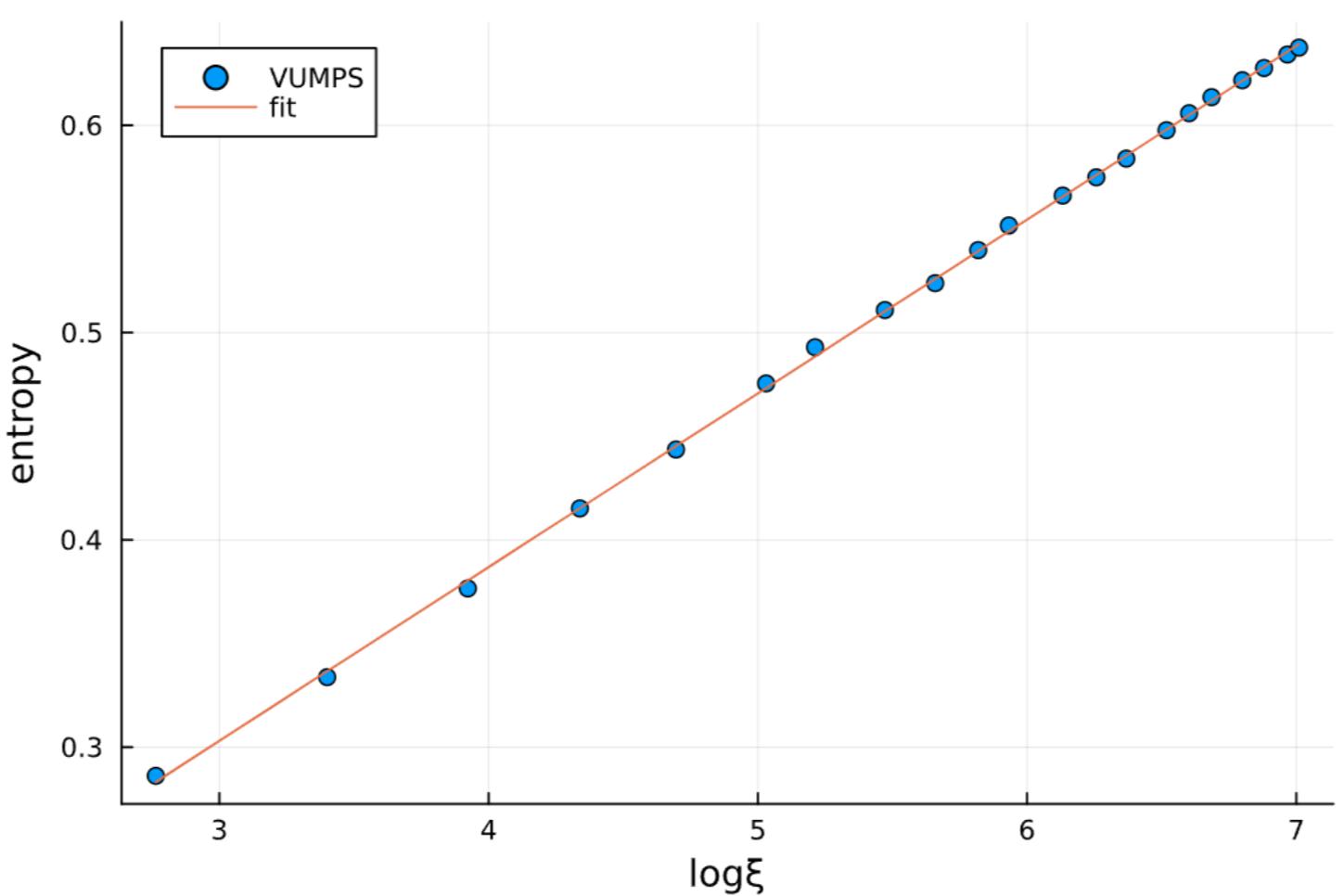}
        \caption{}
        \label{fig:C_3}
    \end{subfigure}
    \caption{(a) The entanglement entropy plotted along the arc $DB$. The horizontal axis $l$ is the length of the arc from point $D$ to the point. (b) The entanglement scaling at the point with $C_3$, $\xi$ is the effective correlation length of iMPS. Fit result is $c=0.503$.}
    \label{fig:DE}
\end{figure}

The points $C_1,C_2,C_3$ and $C_4$ are the phase transition points between neighboring gapped phases. They're located at the quadrisection points of the arc $DE$ and $DB$ on the unit sphere. That's to say all of the following arcs have length $\pi/8$: $DC_1, C_1A, AC_2, C_2E, DC_3, C_3F, FC_4, C_4B$. On the $xOy$ plane, the coordinates of these points are easily calculated: $C_1$ and $C_2$ have coordinates $(0,\pm(\sqrt{2}-1))$, $C_3$ and $C_4$ have coordinates $(2\pm\sqrt{2}, 1)$. The entanglement scaling results show that all of the four critical points have central charge $c=1/2$. In Fig.\ref{fig:DE} we show the entanglement entropy along the arc $DB$ and the entanglement scaling at the point $C_3$. The entropy plot has a clear peak at the point $C_3$ and $C_5$. The entanglement entropy at $D$ and $F$ is zero because they're the topological fixed points\cite{cft_from_tqft}.

\subsubsection{Along the KW line $DG$}
\begin{figure}
    \centering
    \includegraphics[width=0.8\textwidth]{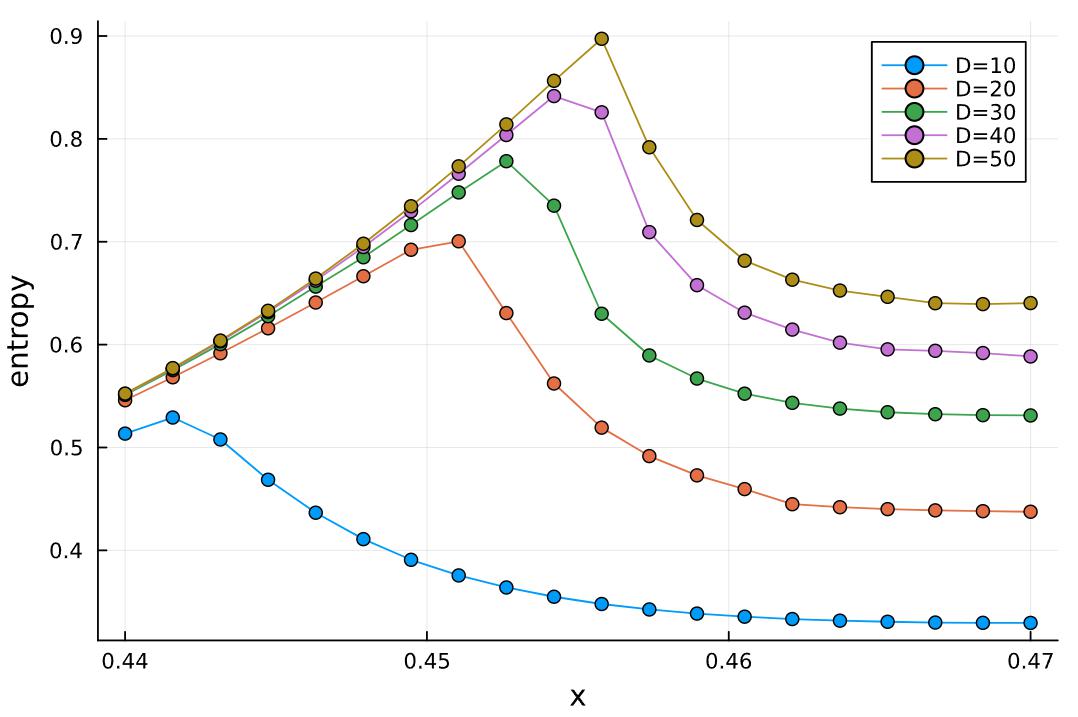}
    \caption{At the vicinity of the point $T_1$, the entanglement entropy is plotted along the line $DG$ for varying iMPS bond dimension. The horizontal axis is the $x$-coordinate of the point. The peak of the entropy is expected to be the tricritical point $T_1$.}
    \label{fig:T1_vary_D}
\end{figure}

On the $xOy$ plane, the KW line $DG$ has equation $y=-\sqrt{2}x+1$. The point $C_5$ and $T_2$ have coordinates $(\sqrt{2}/3,1/3)$ and $(\sqrt{2}/2,0)$ respectively. The entanglement scaling results suggest that both points are CFT points with central charge $c=1$.
The precise location of the point $T_1$ is currently unclear. We provide two methods to estimate the $x$-coordinate of $T_1$, and the two methods give consistent results. 
The first method is to plot the entanglement entropy along the line $DG$, the tricritical point $T_1$ should be the point where the entropy assumes a local maximum. We do this for varying iMPS bond dimension as in Fig.\ref{fig:T1_vary_D}, and extrapolate that the $x$-coordinate of the peak for large bond dimension.
The second method is to test entanglement scaling at the points with $x=0.45$ and $x=0.46$ respectively. The entropy-$\log\xi$ plot shows a clear linear behavior at $x=0.46$ but not at $x=0.45$, suggesting that the system is still in the gapped phase at $x=0.45$ but has already reached the critical line at $x=0.46$. The fit result at $x=0.46$ is $c=1.08$. 
The two methods give consistent results that the $x$-coordinate of $T_1$ is in the range $0.45< x(T_1) <0.46$.

% \subsubsection{Along the line $AB$}
\subsubsection{Along the line $AG$ and $FG$}
\begin{figure}[h]
    \centering
    \begin{subfigure}{0.3\textwidth}
        \centering
        \includegraphics[width=\textwidth]{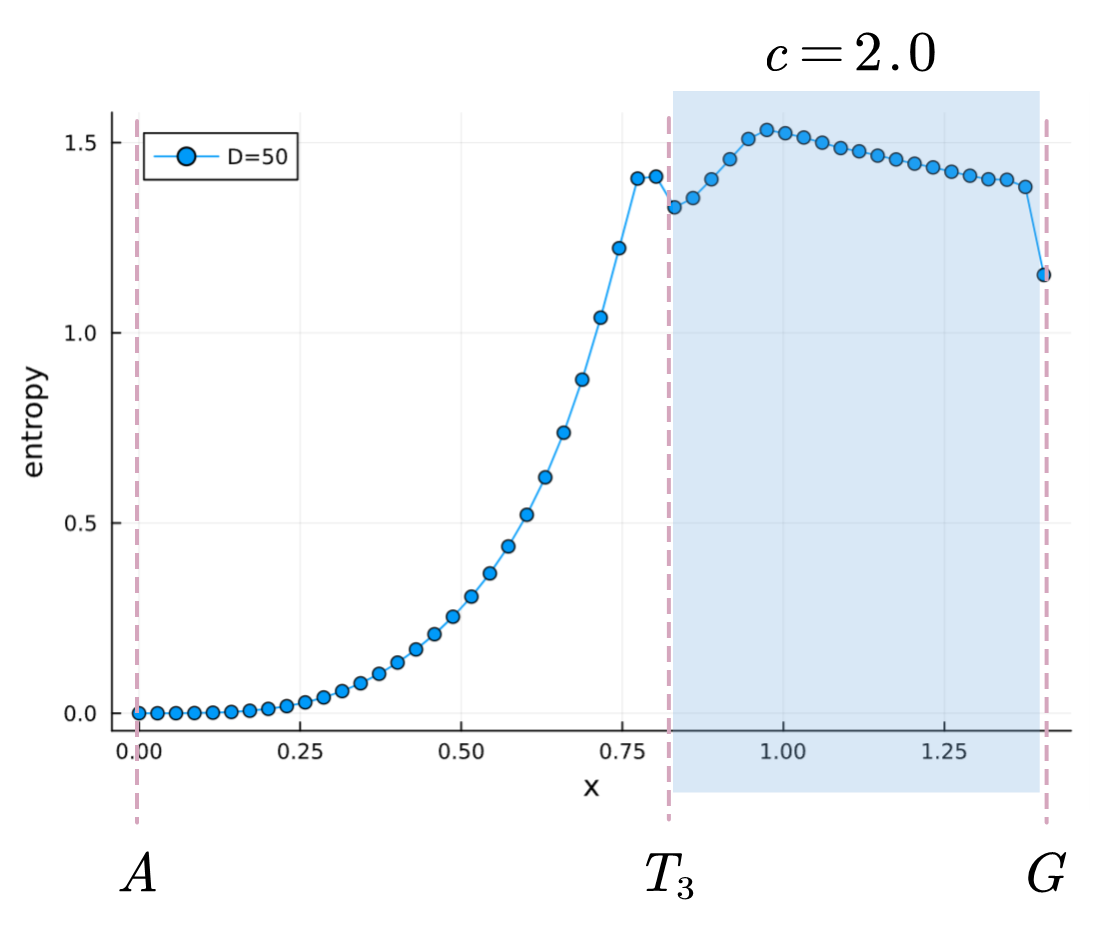}
        \caption{}
        \label{fig:AG_line}
    \end{subfigure}
    \hfill
    \begin{subfigure}{0.3\textwidth}
        \centering
        \includegraphics[width=\textwidth]{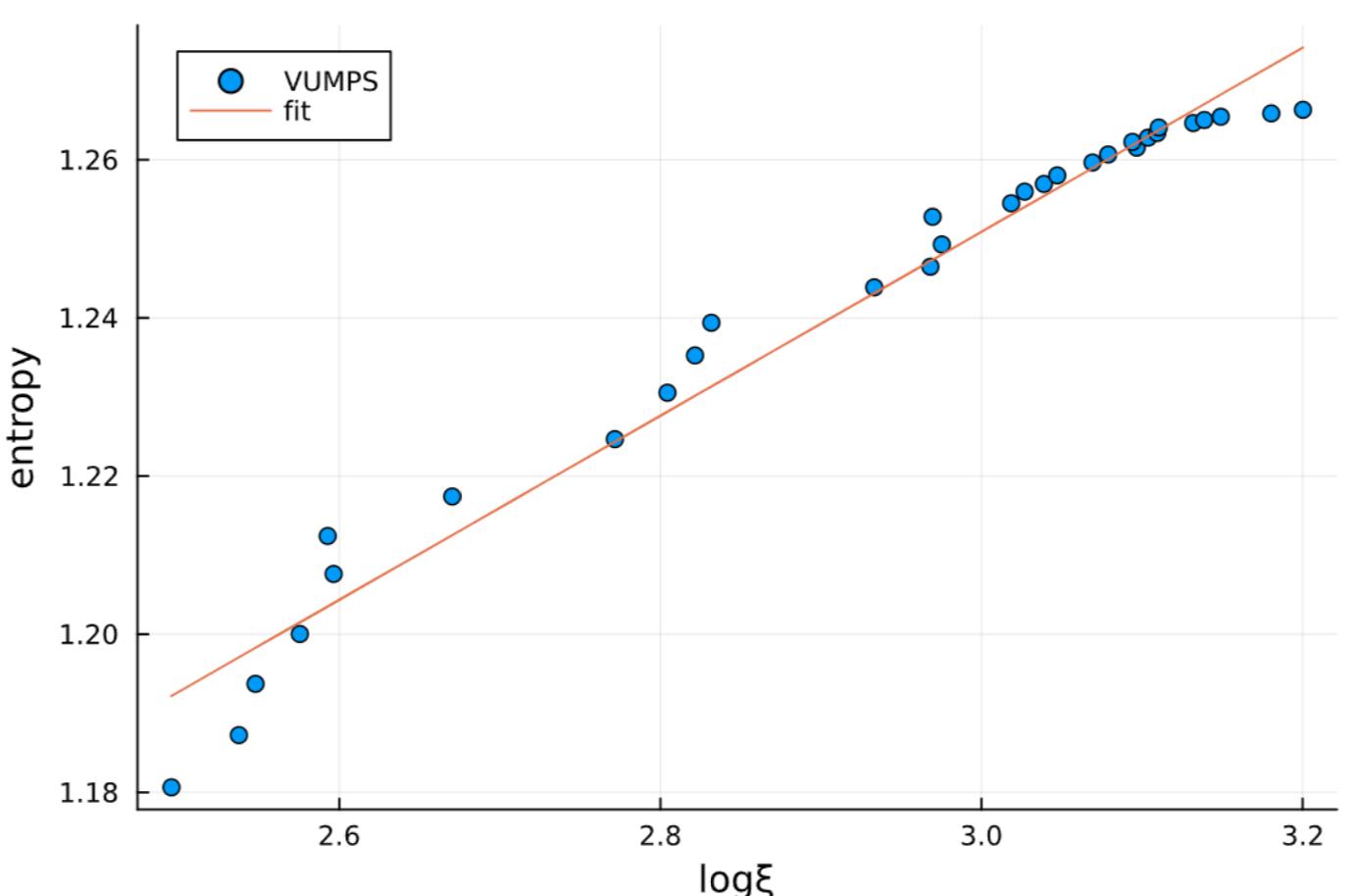}
        \caption{}
        \label{fig:AG_1}
    \end{subfigure}
    \begin{subfigure}{0.3\textwidth}
        \centering
        \includegraphics[width=\textwidth]{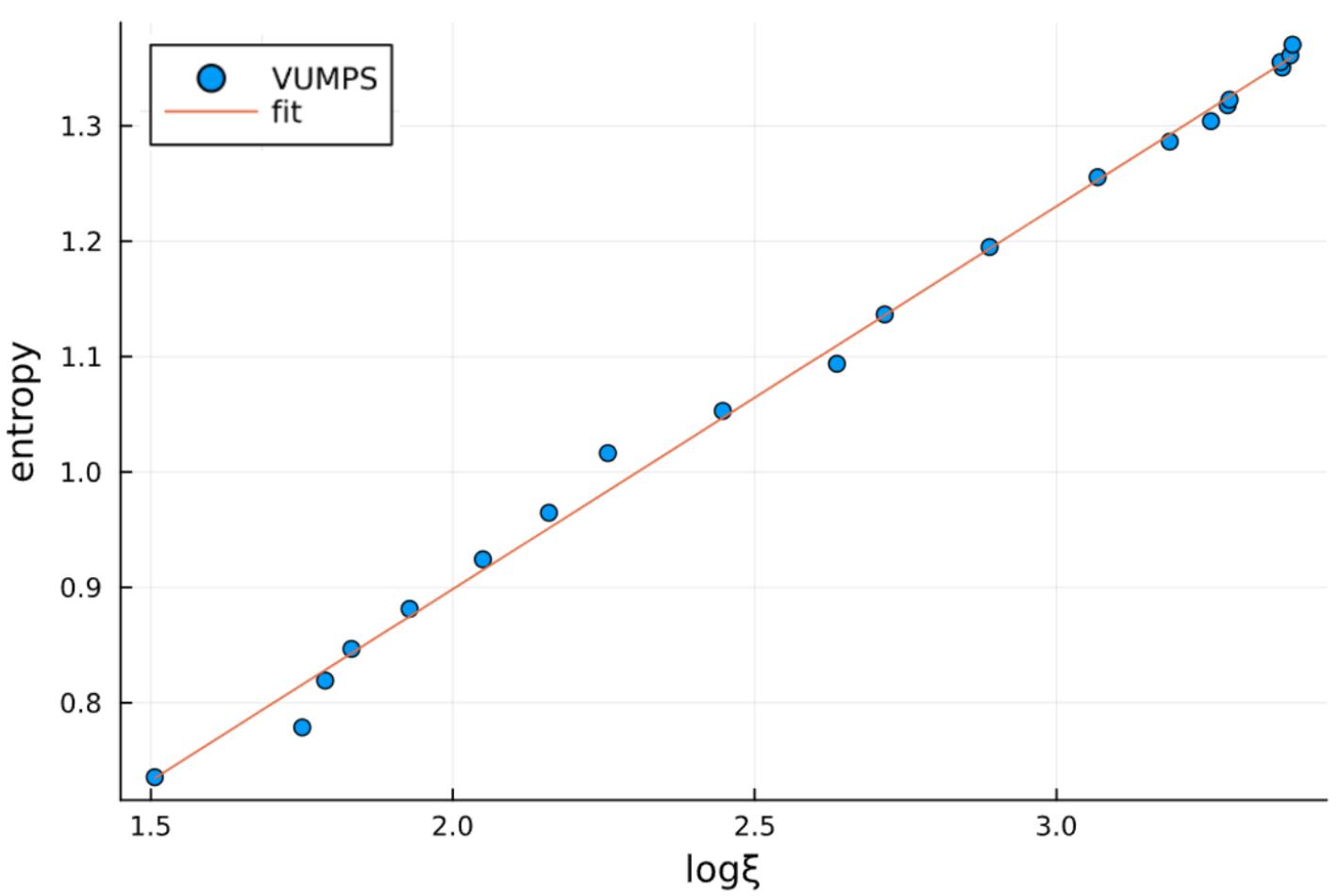}
        \caption{}
        \label{fig:AG_2}
    \end{subfigure}
    \caption{(a) The entanglement entropy plotted along the line $AG$. The horizontal axis is the $x$-coordinate of the point on the line. (b) The entanglement scaling results at the point with $x=0.75$. (c) The entanglement scaling results at the point with $x=0.8$. The fit result is $c=1.99$.}
    \label{fig:AG}
\end{figure}

On the $xOy$ plane, the line $AG$ has equation $y=-1/\sqrt{2}x$ and the line $FG$ has equation $x=\sqrt{2}$. The point $T_3$ and $T_4$ are tricritical points between the CFT phase and two neighboring gapped phases. They are located on the lines $AG$ and $FG$ respectively. $T_3G$ and $T_4G$ are critical lines.  The central charge of these two phase transition lines is $c=2$. The exact location of the points $T_3$ and $T_4$ remain unknown.
Fig.\ref{fig:AG_line} shows the entanglement entropy along the line $AG$ when the iMPS bond dimension is taken to be $D=50$. The entanglement entropy grows as the system approaches $T_3$. After crossing $T_3$, the line $T_3G$ is the phase transition line on which every point is a CFT point with central charge $c=2$. 

Numerical analysis shows that the $x$-coordinate of $T_3$ is in the range $0.75< x(T_3) <0.8$, as can be seen from Fig.\ref{fig:AG_1} and Fig.\ref{fig:AG_2}. The entropy-$\log\xi$ plot deviates from a straight line when $x=0.75$ (and $y=-1/\sqrt{2}x$), which means the entropy tends to saturate for large bond dimension, indicating that the system is still in the gapped phase. However, when $x=0.8$ (and $y=-1/\sqrt{2}x$), the entropy-$\log\xi$ plot shows a clear linear behavior, suggesting that the system has already reached the critical line $T_3G$. At this point the fit result is central charge $c=1.99$.
Using a similar strategy, we also narrow down the range of $y$-coordinate of $T_4$ to be $-0.5< y(T_4)<-0.4$. The entanglement scaling at the point $(\sqrt{2}, -0.5)$ shows a central charge $c=2.01$. We have tested many other points along the line $T_3G$ and $T_4G$ and all of them are consistent with central charge $c=2$.
Thus we conclude that the phase transition lines $T_3G$ and $T_4G$ both have central charge $c=2$.

\subsection{The CFT phase}
\begin{figure}
    \centering
    \begin{subfigure}{0.48\textwidth}
        \centering
        \includegraphics[width=\textwidth]{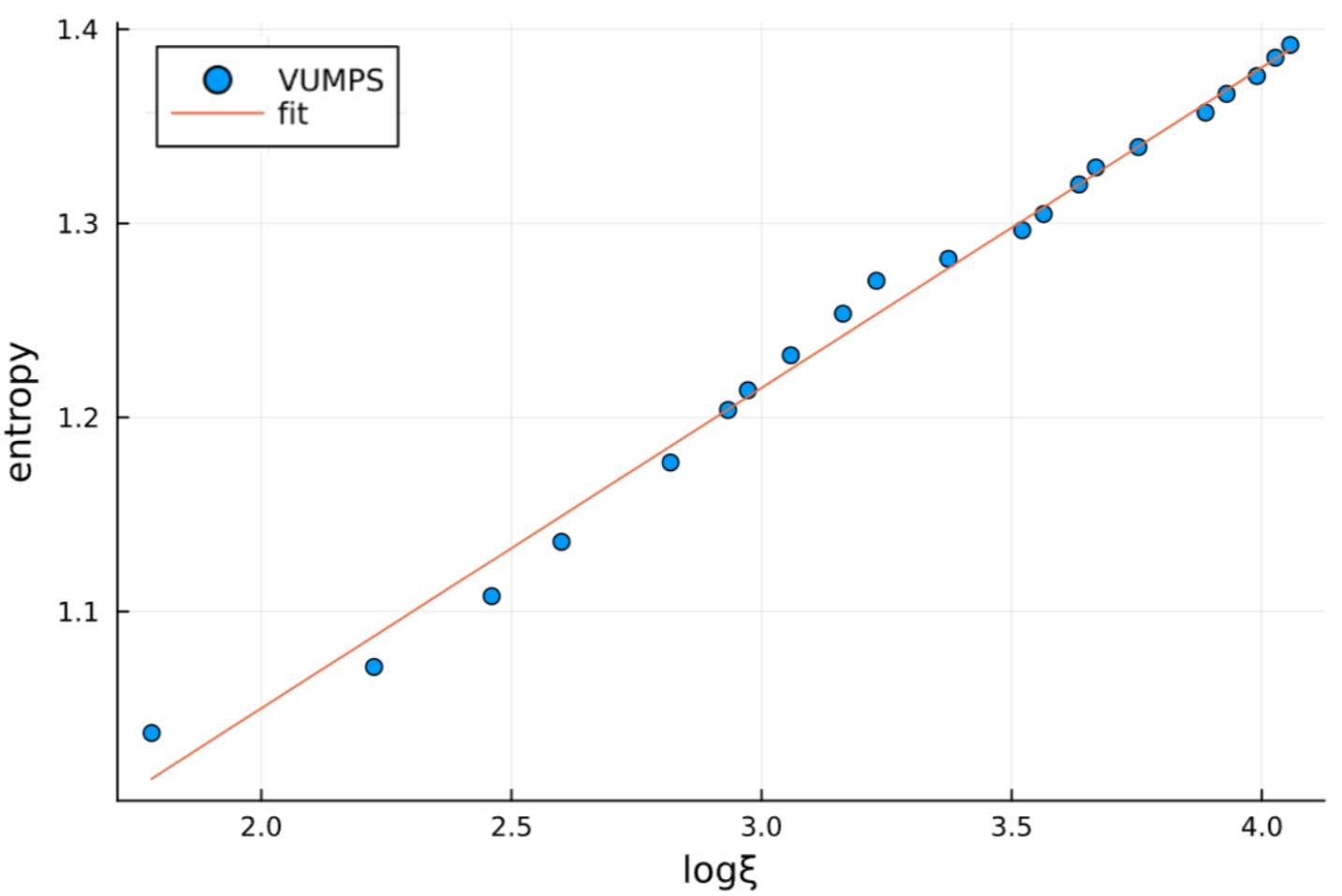}
        \caption{}
        \label{fig:CFT_phase_1}
    \end{subfigure}
    \hfill
    \begin{subfigure}{0.48\textwidth}
        \centering
        \includegraphics[width=\textwidth]{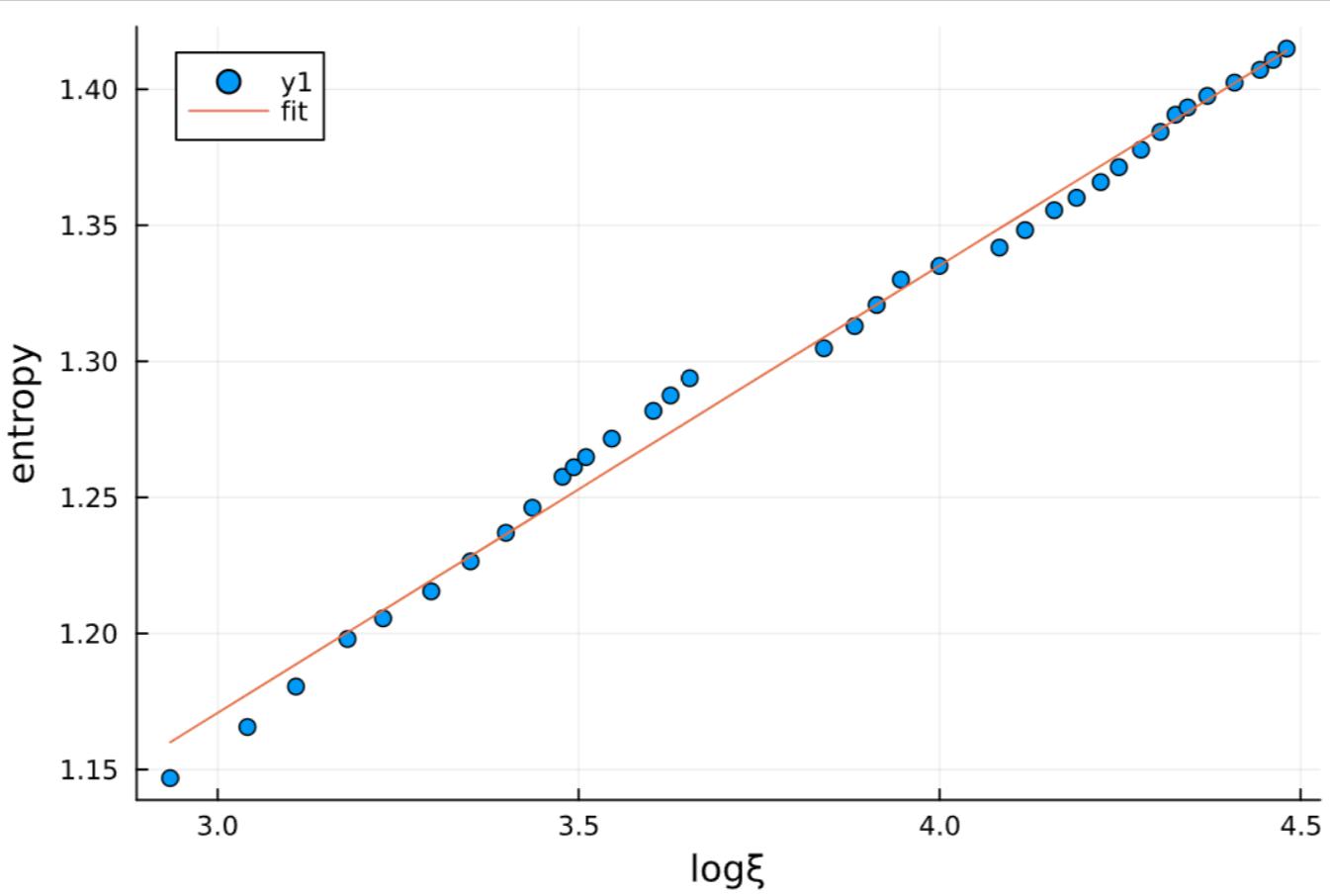}
        \caption{}
        \label{fig:CFT_phase_2}
    \end{subfigure}
    \caption{(a) At the point $(x=0.8, y=-x/\sqrt{2}+0.1)$ fit result shows a central charge $c=0.99$. (b) At the point $(1.2,-0.5)$ fit result shows a central charge $c=0.98$.}
    \label{fig:CFT_phase}
\end{figure}

In the phase diagram Fig.\ref{fig:spherical_plot}, the red region roughly enclosed by $T_1, T_3, T_4, G$ has nearly maximal entropy, indicating that it is a gapless phase. We give numerical evidence that this is indeed a CFT phase with central charge $c=1$. To this end, we have tested entanglement scaling at several points in this region. And at every point we've seen a linear entropy-$\log\xi$ behavior, and the results are consistent with the central charge $c=1$. For example, the entanglement scaling at the point $(x=0.8, y=-x/\sqrt{2}+0.1)$ and $(x=1.2, y=-0.5)$ is shown in Fig.\ref{fig:CFT_phase}. The first point is deliberately chosen to be close to the point $(x=0.8, y=-x/\sqrt{2})$ which is on the phase transition line $T_3G$ with a central charge $c=2$ (see Fig.\ref{fig:AG_2}). We conclude that the red region is a CFT phase with central charge $c=1$.

However, in the CFT phase, only the phase boundaries $T_3G$ and $T_4G$ are distinctly visible and can be clearly identified in the phase diagram. The remaining phase boundaries of the CFT phase are not well resolved at the current level of numerical accuracy and require further investigation.

\section{Conclusion}
In this work, we have systematically investigated the interplay between topological order and criticality within the framework of the strange correlator. By employing the Levin-Wen model with an input fusion category $SU(2)_4$, we constructed a tensor network representation of the strange correlator. This representation is obtained by taking the inner product with another direct product state $\ket{\Omega}$. The state $\ket{\Omega}$ contains two tuning parameters, which allow for the interpolation between gapped topological phases and critical regimes described by CFTs.

Our numerical analysis, based on entanglement entropy computed via iMPS, yielded a detailed phase diagram on the parametric sphere.The results reveal several noteworthy features:
\begin{itemize}
    \item \textbf{Phase Diagram on the Sphere:} A detailed mapping of the phase diagram on unit sphere, highlighting distinct regions of gapped and critical behavior.
    \item \textbf{New Gapped and $c=1$ CFT Phases:} Previously unidentified gapped phases were discovered, along with critical phases described by a central charge $c=1$.
    \item \textbf{Phase Transition Lines with $c=2$:} Transitions between critical and gapped phases occur along lines characterized by a central charge of $c=2$.
    \item \textbf{Explicitly Identified Critical Points:} Several critical points were located precisely, with their central charge values extracted from numerical analysis.
\end{itemize}

These findings underscore the versatility of the strange correlator as a tool for bridging topological quantum field theories and critical phenomena.
% The distinct numerical signatures—logarithmic scaling of entanglement entropy in the critical regime and its saturation in the gapped phases—provide robust evidence for the coexistence and mutual influence of topological order and criticality.
Future research directions include extending this framework to other fusion categories and investigating the interplay between topological phases and critical behavior. Additionally, refining numerical techniques or developing complementary analytical approaches could provide deeper insights into the mechanisms governing the transitions between topological and conformal phases in two-dimensional quantum systems.

In conclusion, our study reinforces the profound connection between topologically ordered states and conformal field theories while opening promising avenues for further exploration in quantum many-body physics.

\appendix
\section{$SU(2)_4$ fusion category data} \label{app:su2_4}
The $SU(2)_4$ fusion category arises from the representation theory of the affine Lie algebra $\widehat{\mathfrak{su}}(2)_4$ and serves as an example of a fusion category. Below, we provide a concise introduction to its fundamental structures.

The $SU(2)_4$ fusion category consists of five simple objects, conventionally labeled as:
\begin{equation*}
    \mathbf{0}, \mathbf{1}, \mathbf{2}, \mathbf{3}, \mathbf{4}
\end{equation*}
where $\mathbf{0}$ is the identity object.

The quantum dimensions of the simple objects are given by $d_{\mathbf{0}}=d_{\mathbf{4}}=1$, $d_{\mathbf{1}}=d_{\mathbf{3}}=\sqrt{3}$, $d_{\mathbf{2}}=2$. The fusion rules are given by
\begin{equation*}
    \begin{array}{llll}
    \mathbf{1} \otimes \mathbf{1}=\mathbf{0}+\mathbf{2} & & & \\
    \mathbf{1} \otimes \mathbf{2}=\mathbf{1}+\mathbf{3} & \mathbf{2} \otimes \mathbf{2}=\mathbf{0}+\mathbf{2}+\mathbf{4} & & \\ 
    \mathbf{1} \otimes \mathbf{3}=\mathbf{2}+\mathbf{4} & \mathbf{2} \otimes \mathbf{3}=\mathbf{1}+\mathbf{3} & \mathbf{3} \otimes \mathbf{3}=\mathbf{0}+\mathbf{2} & \\ 
    \mathbf{1} \otimes \mathbf{4}=\mathbf{3} & \mathbf{2} \otimes \mathbf{4}=\mathbf{2} & \mathbf{3} \otimes \mathbf{4}=\mathbf{1} & \mathbf{4} \otimes \mathbf{4}=\mathbf{0}
\end{array}
\end{equation*}

For a complete list of the F-symbols of $SU(2)_4$, we refer the readers to the appendix A of \cite{ZhenghanWang2015SU2}.

\section*{Acknowledgements}
We thank Ling-Yan Hung, Yidun Wan, Kaixin Ji and Yu Zhao for helpful discussions. We would also like to thank Kaixin Ji for crosschecking the F symbols in the program. We are particularly grateful to Ling-Yan Hung for a careful reading of the draft.
This work is supported by the Beijing Institute of Mathematical Sciences and Applications.

\bibliography{ref}
\bibliographystyle{utphys}
\end{document}